\documentclass[fleqn,usenatbib]{mnras}

\usepackage{newtxtext,newtxmath}

\usepackage[T1]{fontenc}
\usepackage{ae,aecompl}
\usepackage{listings}
\usepackage{lipsum} 

\usepackage{graphicx}	
\usepackage{amsmath}	
\usepackage{amssymb}	
\usepackage{color}
\usepackage{cprotect}
\definecolor{linkcolor}{rgb}{0,0,0.25}
\hypersetup{
  colorlinks=true,        
  linkcolor=linkcolor,    
  citecolor=linkcolor,    
  filecolor=linkcolor,    
  urlcolor=linkcolor      
}

\makeatletter
\renewcommand{\@printed}{}
\makeatother

\newcommand{\figurename}{Figure}

\newcommand{\eqnname}{Equation}
\newcommand{\secname}{Section}

\definecolor{darkgreen}{rgb}{0.0, 0.5, 0.0}

\definecolor{darkblue}{rgb}{0.0, 0., 0.7}

\newcommand{\gaia}{\emph{Gaia}}

\title[Vertical waves in \gaia\ DR2]{Vertical waves in the solar neighbourhood in \gaia\ DR2}

\author[Bennett \& Bovy]{
Morgan Bennett\thanks{E-mail: bennett@astro.utoronto.ca}\&
Jo Bovy\thanks{Alfred P. Sloan Fellow}
\\
Department of Astronomy and Astrophysics, University of Toronto, 50 St. George Street, Toronto, Ontario, M5S 3H4, Canada
}

\date{}

\pubyear{2018}

\begin{document}
\label{firstpage}
\pagerange{\pageref{firstpage}--\pageref{lastpage}}
\maketitle

\begin{abstract}
The vertical structure and dynamics of stars in our local Galactic neighbourhood contains much information about the local distribution of visible and dark matter and of perturbations to the Milky Way disc. We use data on the positions and velocities of stars in the solar neighbourhood from \gaia\ DR2 and large spectroscopic surveys to investigate the vertical number counts and mean-velocity trend as a function of distance from the local Galactic mid-plane. We perform a detailed measurement of the wave-like North-South asymmetry in the vertical number counts, which reveals a number of deficits at heights $\approx 0.4\,\mathrm{kpc}$, $\approx 0.9\,\mathrm{kpc}$, and $\approx 1.5\,\mathrm{kpc}$, and peaks at $\approx 0.2\,\mathrm{kpc}$, $\approx 0.7\,\mathrm{kpc}$, and $\approx 1.1\,\mathrm{kpc}$. We find that the asymmetry pattern is independent of colour. The mean vertical velocity is almost constant to $<1\,\mathrm{km\,s}^{-1}$ within a few 100 pc from the mid-plane and then displays a North-South symmetric dip at $\approx0.5\,\mathrm{kpc}$ with an amplitude of $\approx 2\,\mathrm{km\,s}^{-1}$ that is a plausible velocity counterpart to the main number-count dip at a similar height. Thus, with \gaia\ DR2 we confirm at high fidelity that the local Galactic disc is undergoing a wave-like oscillation and a dynamically-consistent observational picture of the perturbed local vertical structure emerges for the first time. We also present the most precise and accurate determination of the Sun's height above the local Galactic mid-plane, correcting for any asymmetry in the vertical density: $z_\odot = 20.8 \pm 0.3\,\mathrm{pc}$.
\end{abstract}





\section{Introduction}

The Milky Way is a complex system, the structure and dynamics of which are still being untangled. Much can be learned about the structure and dynamics of the Milky Way from observations of its vertical structure. Studies of the local vertical structure were pioneered by \citet{oort32} in his study on the vertical force of the Galactic disc. Over the past century, ever more detailed measurements of the vertical structure and kinematics of stars has led to increasingly precise determinations of the local mass distribution and dynamics \citep[e.g.,][]{1984ApJ...276..169B,1989MNRAS.239..605K,2000MNRAS.313..209H,bovydm}. Until recently, a common assumption in these studies is that the dynamics of the local Galactic disc is in equilibrium and combined with large stellar kinematic surveys, this assumption has allowed for precise measurements of the mass distribution in the Milky Way disc \citep[e.g.,][]{2013ApJ...779..115B}. However, recently, clear non-equilibrium effects in the vertical structure in the solar neighbourhood have been observed as a wave-like perturbation in the vertical number counts of stars \citep{widrow12} and as a spiral pattern in the angular-momentum-painted vertical phase-space distribution \citep{antoja18,binney18,darling18}. These both likely result from the dynamical influence of a fly-by of a large satellite galaxy \citep[e.g.,][]{widrow14}. Therefore, equilibrium is no longer an acceptable assumption. For example, accounting for these deviations from equilibrium can have important impacts on measurements of the surface density of the disc and of the local dark-matter density assuming equilibrium \citep{banik17}. We aim to provide a detailed measurement of the form and amplitude of the wave-like oscillations in number counts and vertical velocity as a first step in modelling their impact on the local dynamics. Modelling vertical oscillations observed in the density and velocity caused by dynamical perturbations can tell us about not only the dynamical history of the disc, but also about fundamental properties of the Galaxy itself, because the disc's response depends on its internal structure \citep{2015MNRAS.450..266W}.

The first evidence of a departure from equilibrium in the vertical dynamics in the solar neighbourhood was discovered by \citet{widrow12}. They used the Sloan Digital Sky Survey (SDSS) Sloan Extension for Galactic Understanding and Exploration (SEGUE) data to calculate the asymmetry in the number counts of stars above and below the Galactic mid-plane---we refer to this as the ``north-south asymmetry''. If the disc were in equilibrium, we would expect the density to be distributed symmetrically about the Galactic mid-plane and any deviation from this indicates the presence of a perturbing force and/or incomplete phase mixing. \citet{widrow12} found an underlying wave-like pattern in the number count density of stars. They also found that the asymmetry is independent of colour and that it is therefore a property of the disc itself and not of a single group of stars. Further work on this was done by \citet{yanny13} using the ninth data release of the SDSS. They confirmed the original results of \citet{widrow12} while performing a thorough analysis of the errors resulting from using photometric parallaxes. Little more has been done on measuring the asymmetry in the solar neighbourhood because it requires the selection function of a survey to recover the true number of stars in a volume and therefore the density \citep{bovyselect}. \gaia\ DR2 is particularly useful for these purposes as it contains geometric parallaxes which have significantly less systematic errors than other distance measurement techniques, and it is also complete over a wide range of apparent magnitudes. It therefore provides us a robust sample of stars which we can use to look at the stellar number counts and therefore the density of the Galactic disc. 

The second indicator that the Galactic disc is out of equilibrium and undergoing oscillations is the behavior of the mean vertical velocity with distance from the mid-plane. For an equilibrium distribution, the mean vertical velocity should be zero at all heights. This was also first discussed by \citet{widrow12} using the SDSS SEGUE data and they found evidence of a breathing mode, that is, a north-south asymmetry in the mean velocity. There have since been surveys which have improved distance measurements and allow improvements to the vertical velocity measurement. Recently, \citet{carrillo18} performed a thorough analysis using the Tycho-Gaia Astrometric Solution (TGAS) catalogue and several different distance measurement techniques combined with radial velocities from the Radial Velocity Experiment (RAVE) to examine the mean vertical velocities as a function of vertical height at different radii. They found that the Galactic disc is undergoing a breathing mode interior to the solar radius, while exterior to the solar radius the velocities display a bending mode---a non-vanishing mean vertical velocity that is symmetric with respect to the mid-plane. Finally, with the release of \gaia\ Data Release 2 (DR2), \citet{gaia18a} investigated the kinematics in the solar neighbourhood. They found evidence of a vertical velocity gradient at different radii with amplitudes between 4-6 km s$^{-1}$. The precision of \gaia\ DR2 parallaxes allows for a much simpler analysis of velocities with the accuracy required for detecting small perturbations.

While at this point there is good evidence of a significant perturbation in both the stellar density and stellar kinematics in the solar neighbourhood, the signals detected in these two tracers are not consistent with each other. An asymmetry in the density should correspond to a symmetric velocity perturbation, but no plausible counterpart to the wave structure in the density has been found so far in the vertical kinematics. In this paper, we leverage the power of \gaia\ DR2 to address this.

\gaia\ DR2 provides us with an unprecedented opportunity to examine vertical waves in the solar neighbourhood. Previous studies have had to rely on photometric or spectroscopic parallaxes or other methods of determining distances \citep{widrow12,williams13}. With \gaia\ DR2, we have a large number of stars with 5-dimensional parameters, including geometric parallax, which allows us to investigate the local vertical structure with extraordinary accuracy. As mentioned, \gaia\ is also complete over a wide range of magnitudes which allows us to calculate the number densities without the complication of a selection function \citep{gaia18b}. Finally, with radial velocities for over 7 million stars, \gaia\ DR2 allows us to not only examine the density asymmetry, but also trends in the vertical velocity in the solar neighbourhood to unprecedented levels of precision \citep{gaia18a}. 

In this paper, we begin by outlining the data used for measuring the number count asymmetry and the vertical velocities, as well as what quality cuts were made, in \secname~\ref{sec:data}. \secname~\ref{sec:VC} then examines the number counts above and below the mid-plane for several different colour bins. We outline the analysis necessary to obtain the asymmetry in Section 3.1 and present our measurement of the vertical density asymmetry in Section 3.2. \secname~\ref{sec:zsun} applies the results from the asymmetry analysis to fit the number counts to a model which includes the asymmetry to obtain an accurate estimate of the solar height above the mid-plane. \secname~\ref{sec:vel} discusses the analysis and results of the mean vertical velocity signature in \gaia\ DR2. Finally, in \secname~\ref{sec:dandc} we conclude the paper by comparing our results to those from previous work and discussing the implications of our new measurements of the vertical waves.

\section{Data}\label{sec:data}
\subsection{Number Counts}

\begin{figure}
\includegraphics[width=\columnwidth]{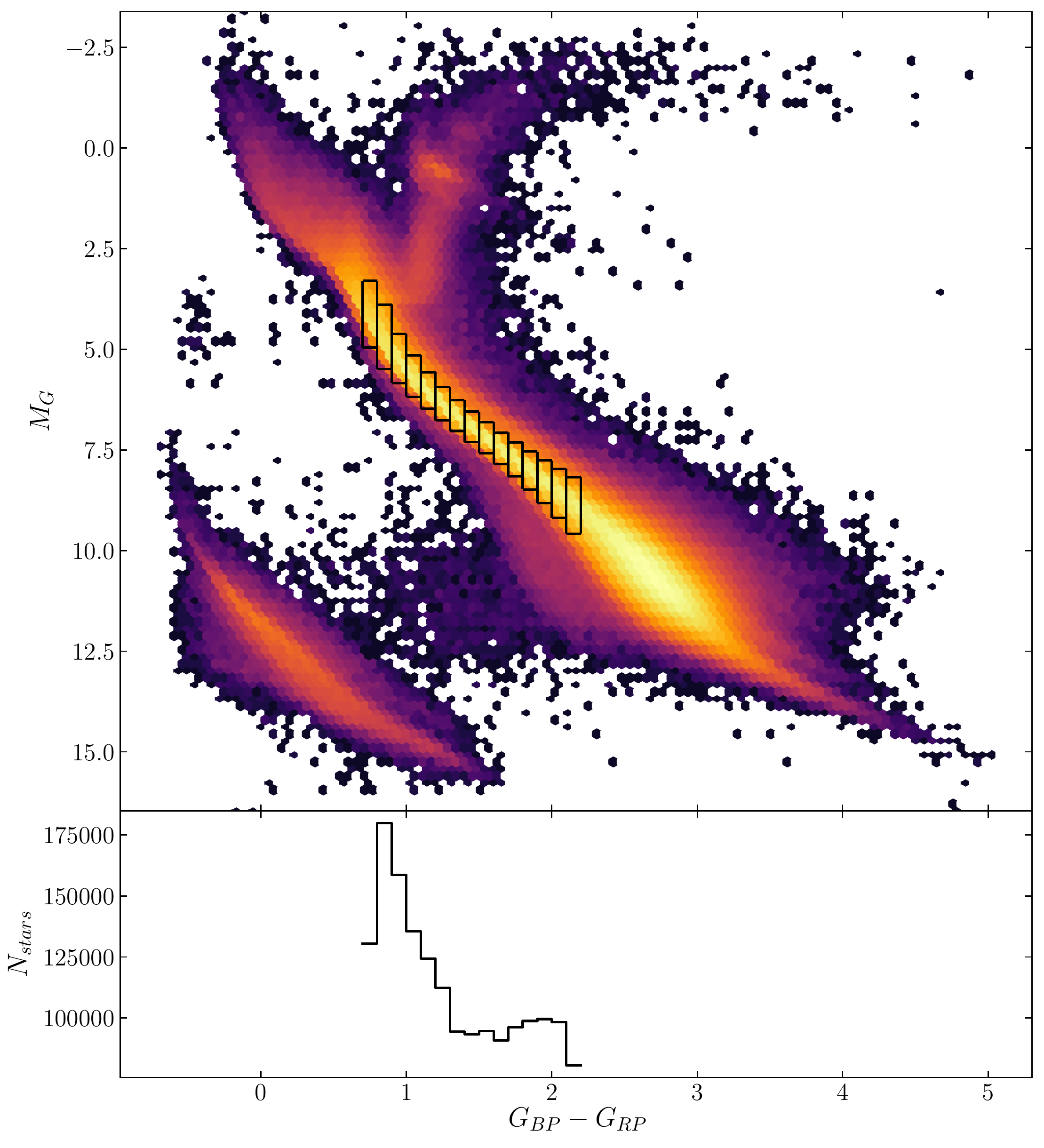}
    \caption{Colour magnitude diagram using the white (G), red (RP), and blue (BP) passbands from \gaia. The boxes outlined in black show the different colour bins we use in the analysis. The bottom panel shows the total number of stars in each bin after both the selections described in \secname~\ref{sec:data} and the completeness cuts described in \secname~\ref{sec:VC} are applied.} 
    \label{fig:cmd}
\end{figure}

\begin{figure*}
	\includegraphics[width=\textwidth]{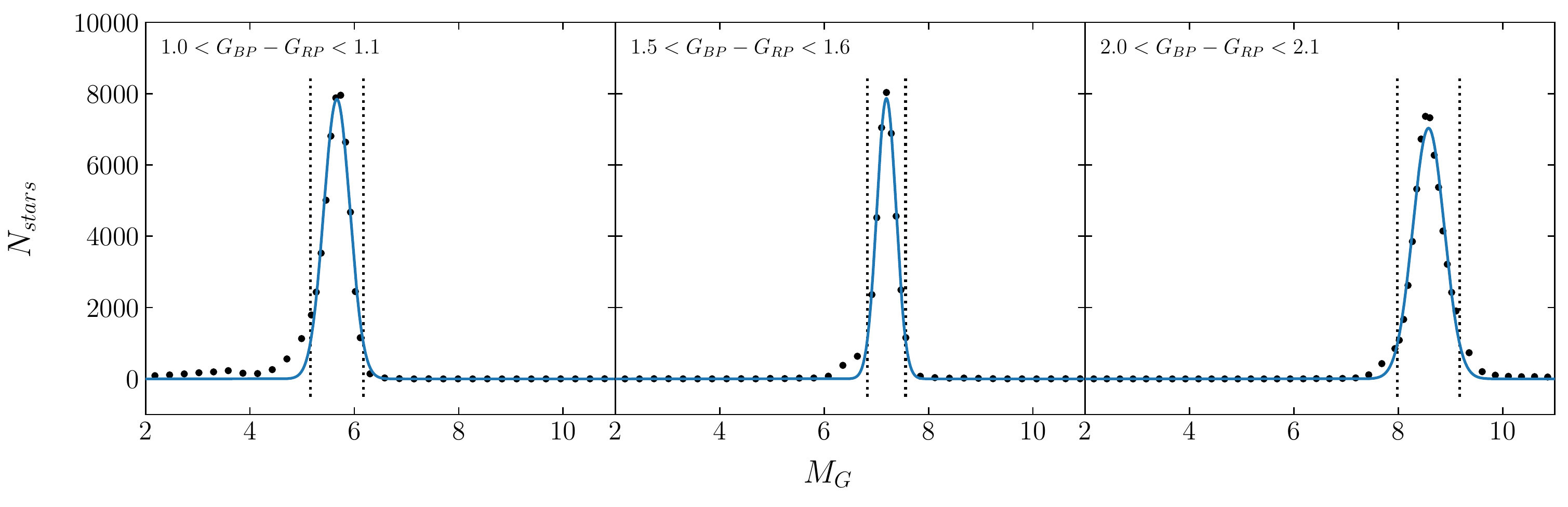}
    \caption{Number counts as a function of absolute magnitude for three different colour bins. The Gaussian fit to the peak of the main sequence is shown by the blue line. The vertical dotted lines show the $2\sigma$ cut-off corresponding to the \texttt{M\_bright} and \texttt{M\_faint}  magnitudes cuts in our query.} 
    \label{fig:fit}
\end{figure*}

The first step in investigating perturbations to the vertical structure of the Milky Way is looking at the stellar number counts above and below the Galactic disc. The first important question in relation to this is how complete \gaia\ DR2 is at different apparent magnitudes. To investigate this, we compare the number counts of stars in \gaia\ DR2 to those in the 2MASS Point Source Catalogue \citep{2006AJ....131.1163S}, which is $> 99\,\%$ complete down to magnitudes of $J = 15.8$ and $K = 14.3$ over almost the entire sky, except for a few regions near the Galactic plane and within a few degree of the Galactic centre. We compare the number counts of stars with $-0.25 < J-K_s < 1.25$ in bins of $\Delta J = 1\,\mathrm{mag}$, which roughly corresponds to $-0.65 < G_{\mathrm{BP}}-G_{\mathrm{RP}} < 2.55$ and thus encompasses the colour range that we consider below. Using the official \gaia\ DR2--2MASS cross-match best-neighbour table  \citep{gaiadr2crossmatch}, we find that typically $\approx99\,\%$ of stars in 2MASS are contained within the cross-match. Checking the completeness of the cross match itself by computing the relative number of stars in \gaia\ DR2 that are contained in the 2MASS cross match, we find that the match completeness is also $\approx99\,\%$. Dividing the \gaia--2MASS completeness by the match completeness in small regions of the sky, we find that this equals $100\,\%$ with an uncertainty of $<0.1\,\%$ down to $J=11$. At fainter magnitudes, \gaia\ contains slightly more stars than 2MASS in exactly the locations on the sky identified as having low completeness in the 2MASS documentation (primarily near the Galactic centre). Thus, we conclude that \gaia\ DR2 is complete down to the 2MASS completeness limit of $J = 15.8$. For typical stars that we consider below, $G-J \approx 1.3$ and for these stars \gaia\ DR2 is therefore complete down to at least $G = 17$. Because Gaia's sky mapper CCD which detects stars that end up in the catalogue uses the broad G-band filter, a detection or non-detection of a star depends only on the magnitude of the star in the G-band, not on the colour of the star. Therefore, we expect the \gaia\ completeness to be independent of colour. This completeness limit in fact applies to all types of stars. We have checked the completeness up to a bright magnitude of $J=3$, leading to the conclusion that \gaia\ DR2 is complete over at least the range $3 \lesssim G \lesssim 17$. In what follows, we select stars with a more conservative limit of $7 < G < 17$. We do this because, while \gaia\ appears to be complete when comparing to 2MASS, \citet{gaia18b} explicitly reports that the survey is incomplete at $G < 7$. Because there are not many stars or Galactic volume in the magnitude range between 3 and 7, we choose not to include this magnitude range so that we can be confident in the completeness. 

When calculating the number count, we select main sequence stars in cylinders centered on the Sun with radii of 250 pc that are perpendicular to the disc's mid-plane. We choose to use main sequence stars for two reasons. First, given the apparent magnitudes at which \gaia\ DR2 is complete, main sequence stars have absolute magnitudes in a range that allows us to probe both close to the disc ($\sim$10 pc) and further away ($\sim$2 kpc). Second, and most importantly, they trace the density of the disc with little dependence on the star-formation history and therefore trace the total stellar mass. 

Because \gaia\ DR2 is complete over a range of apparent magnitudes and there is a strong relation between colour and absolute magnitude, it is easier to define volumes over which \gaia\ DR2 is complete for stars in narrow colour ranges than in one broad colour bin. Therefore, we split our sample by colour. Additionally, if we measure an asymmetry in the vertical number counts that is the same in all colour bins, we can be confident that we are seeing a true, disc-dynamical effect, rather than the result of a systematic error or of a perturbation in a small subset of stars. To select the main sequence for each colour bin, we need to make cuts on the absolute magnitude of stars. We want to ensure that we are looking at roughly the same percentage of main sequence stars for each colour bin when we make these cuts. To do this, we first look at a test case where we take stars in a sphere with a radius of 250 pc centered on the Sun and plot their colour magnitude diagram as shown in \figurename~\ref{fig:cmd}. These stars also have additional photometric and astrometric cuts on error described by \citet{gaiacmd} in Appendix B. Next, for each colour bin we calculate the number of stars at each magnitude and fit a Gaussian to the main sequence peak. This allows us to remove objects like white dwarfs and giants from our selection and ensure that we are selecting a similar fraction of main-sequence stars for each colour bin. \figurename~\ref{fig:fit} shows the results of this fit for three example colour bins spaced throughout our entire colour range. In both the first and second plot, we clearly see the ability of the fit to remove giants from our selection. The magnitude cuts we choose are defined as the $2\sigma$ intervals of this Gaussian fit and are shown as vertical dotted lines in \figurename~\ref{fig:fit}. The resulting colour--absolute-magnitude selections are shown as boxes in \figurename~\ref{fig:cmd}.

We then use these bins to select our colour and absolute magnitude cuts when querying the \gaia\ archive. We also make quality cuts on parallax since we estimate the distances as 1/parallax. This cut does effect the completeness of our sample further out. The median uncertainties in parallax for a sample of stars in a cylinder with a radius of 250 pc and apparent magnitudes between 16.5 and 17 is approximately 0.095 mas. Since we cut on 20\% error in parallax, that means this cut restricts us to a cylinder approximately 2 kpc in height. The final query that results from these cuts is given by:

\begin{lstlisting}
SELECT (1/parallax)*cos(RADIANS(b))*cos(RADIANS(l)) AS x, 
cos(RADIANS(b))*(1/parallax)*sin(RADIANS(l)) AS y, 
(1/parallax)*sin(RADIANS(b)) AS z
FROM gaiadr2.gaia_source
WHERE parallax > 0
AND parallax_over_error > 5
AND (1/parallax)*cos(RADIANS(b)) < 0.25
AND phot_g_mean_mag BETWEEN 7 AND 17 
AND bp_rp BETWEEN br_min AND br_max
AND -5*LOG(1000/parallax)/LOG(10)+5+phot_g_mean_mag BETWEEN M_bright AND M_faint
\end{lstlisting}

We choose to not make cuts on the quality of the astrometric values of the sample recommended by \citet{gaiacmd}. A test of the effect of those quality cuts on our analysis showed that they do not have a noticeable effect on the main conclusions of the paper, but removed approximately 10\% of stars from each colour bin. Our focus is on completeness of the sample, so we chose not to include these quality cuts and instead keep as many stars as possible. 

The variables \texttt{br\_min} and \texttt{br\_max} are the red and blue cuts on colour for each bin, respectively, and are shown as the vertical black lines in \figurename~\ref{fig:cmd}. \texttt{M\_bright} and \texttt{M\_faint} are the minimum and maximum absolute magnitude of each colour bin, indicated by the horizontal black lines in \figurename~\ref{fig:cmd}. Our raw data contains a total of 1,963,018 stars. However, in \secname~\ref{sec:VC} we discuss further cuts made in the analysis related to the volume over which each colour range is complete. The histogram in the bottom panel of \figurename~\ref{fig:cmd} represents the number of stars in each colour bin after we have also done a cut on distance described in \secname~\ref{sec:VC}. 

\subsection{Velocities}\label{sec:data_vel}
\begin{figure*}
\includegraphics[width=2\columnwidth]{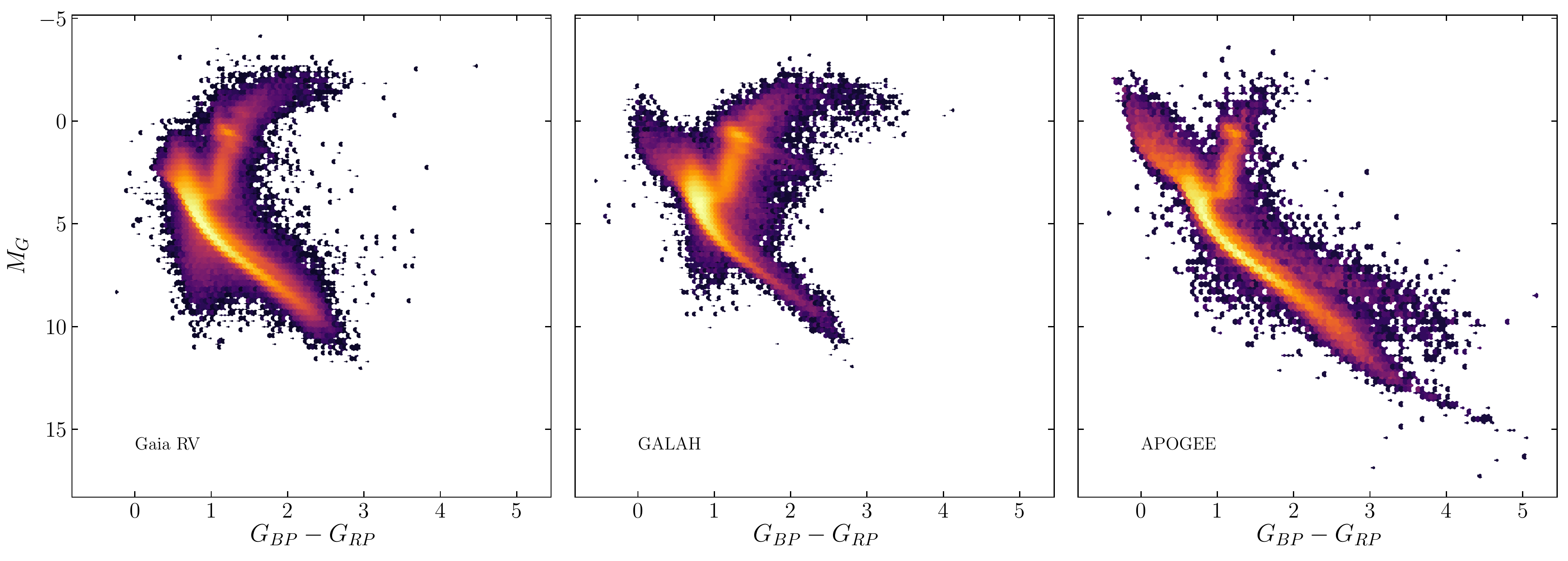}
    \caption{Colour-magnitude diagrams for the three different radial velocity samples used to examine the mean vertical velocity trends. Beyond the query discussed in \secname~\ref{sec:data}, these stars have also undergone a distance cut outlined in \secname~\ref{sec:vel}.} 
    \label{fig:cmd_vel}
\end{figure*}
We want to look at not only the departure from equilibrium in the vertical density of the Milky Way, but also the signature in the vertical velocities. To do this, we use the radial velocity sample from \gaia\ DR2 which includes over 7 million stars \citep{gaia18b}. As a comparison, we also use radial velocity data from the APO Galactic Evolution Experiment (APOGEE) and the GALactic Archaeology with HERMES (GALAH) survey, which allows us to compare velocity trends across different surveys. To estimate distances to stars in these surveys, we match to the entire \gaia\ DR2 catalogue. To download and match the APOGEE and GALAH data to \gaia\ DR2, we use the \texttt{gaia\_tools} Python package\footnote{\url{https://github.com/jobovy/gaia_tools}} \citep{bovyselect}.

None of the surveys we use have a selection cut on velocity, therefore it is accurate to assume the radial velocity sample provides unbiased kinematics at any position. This means that there are no preferred velocities in the sample and our data sets are therefore an accurate sampling of velocities of all stars at a given location in the Milky Way. For radial velocities, like the number count sample, we cut on the error in parallax in order to obtain accurate distances. This is the only cut we make to the sample to keep it as complete as possible. 
\begin{lstlisting}
SELECT radial_velocity, radial_velocity_error,
ra, dec, parallax, parallax_error, pmra, pmra_error, pmdec, pmdec_error
FROM gaiadr2.gaia_source
WHERE parallax IS NOT Null AND parallax_over_error>5.
AND radial_velocity IS NOT Null
\end{lstlisting}

Radial velocities require brighter stars and, to sample the same volume as covered by our number-counts sample described above, therefore need intrinsically brighter stars. Thus,  all three of the radial velocity surveys have $G_{BP}-G_{RP}$ colours comparable with the bluer bins from our number count sample. The exact location of the samples in colour and absolute magnitude are shown in \figurename~\ref{fig:cmd_vel}. Unlike the number counts, we plot the colour and magnitude of the stars actually in our sample after all cuts have been made and not an example subset. The radial velocity samples also include a fair number of giants which have evolved off the main sequence. The \gaia\ RV sample has over 6 million stars and the GALAH and APOGEE sample each have approximately 300,000 stars. After the distance cuts described in \secname~\ref{sec:vel}, the \gaia\ RV sample has a total of 864,268 stars, GALAH contains 160,686 stars, and the APOGEE sample consists of 85,234 stars. 

\section{Vertical counts in \gaia\ DR2}\label{sec:VC}
\subsection{Analysis}

Before calculating the asymmetry, there are a few corrections to the data which need to be made. First, we know the apparent magnitudes over which the samples are complete, but we need to consider what physical distances correspond to these magnitudes. Second, we need to determine the vertical height of the Sun above the mid-plane, $z_\odot$. This is necessary, because the raw vertical distances are measured with respect to the Sun and we need to correct these to be with respect to the mid-plane before computing the asymmetry. 

The height of our cylinders over which each colour sample is complete is determined by the apparent magnitudes at which \gaia\ DR2 is complete and the absolute-magnitude range of the colour sample. Using these, we can use the distance modulus to calculate distances over which the cylinder is complete. The bright limit (apparent magnitude of 7 and \texttt{M\_bright}) tells us how close to the plane our sample is complete, $d_1$. The faint limit (apparent magnitude of 17 and \texttt{M\_faint}) tells us the furthest distance, $d_2$, at which the cylinder is complete. 
\begin{equation*}
d_1= 10^{\frac{7-\mathtt{M\_bright}}{5}-2}\hspace{30pt}d_2= 10^{\frac{17-\mathtt{M\_faint}}{5}-2}
\end{equation*}
This is not equivalent to the height at which the cylinder is complete, because we also have to account for the radius of the cylinder. This means that for a given colour bin, the maximum height at which our colour bin is complete is actually given by 
\begin{equation}
h= \sqrt{{d_2}^2-(0.25 \textrm{ kpc})^2}
\end{equation} 
In our analysis, we take the minimum of $h$ and 2 kpc, for reasons outlined in \secname~\ref{sec:data}. The number of stars included in each colour bin after these cuts on distance are included in the bottom panel of \figurename~\ref{fig:cmd}. The total number of stars across all colour bins is approximately 1.8 million stars. 

The fact that the Sun is not perfectly located in the Galactic mid-plane manifests as a shift in the number counts, an effect that needs to be corrected before calculating the asymmetry in the number counts. To do this, we fit a two component model to number counts in bins of width $\Delta z = 25$ pc:
\begin{equation}
N(z_{\mathrm{obs}}) = N_0\left(\textrm{sech}^2\left(\frac{z_{\mathrm{obs}}+z_\odot}{2H_1}\right)+f\textrm{sech}^2\left(\frac{z_{\mathrm{obs}}+z_\odot}{2H_2}\right)\right)
\label{eq:ncount}
\end{equation}
Where $z_{\mathrm{obs}}$ is the heliocentric vertical height of each vertical cylindrical bin, $N_0$ is simply a scaling factor, $z_\odot$ is the vertical position of the Sun, $f$ describes the relative importance of each component, and finally $H_1$ and $H_2$ are roughly equivalent to scale heights of each component.

Since we are dealing with number counts and some of the bins in $z$ contain on the order of 10 stars, the uncertainty in our data is described not by a Gaussian distribution, but by a Poisson distribution. This means that the likelihood is given by:
\begin{align}
\ln p(N_\mathrm{obs}|N)=\sum_i \left[-N_i+N_{\mathrm{obs},i}\ln(N_i)-\ln(N_{\mathrm{obs},i}!)\right],
\end{align}
where $N_i$ is shorthand for $N(z_i)$ from \eqnname~\ref{eq:ncount}, $N_{\mathrm{obs},i}$ is $N_{\mathrm{obs}}(z_i)$ and is the true observed number count and $z_i$ is simply the midpoint of the different bins in vertical position. The last term is independent of the models and simply adds a constant term to the likelihood for all models and is therefore ignored in the maximum likelihood estimator. For the priors, we assume a uniform distribution between reasonable values for the parameters of \eqnname~\ref{eq:ncount}. 
\begin{figure*}
\includegraphics[width=1.9\columnwidth]{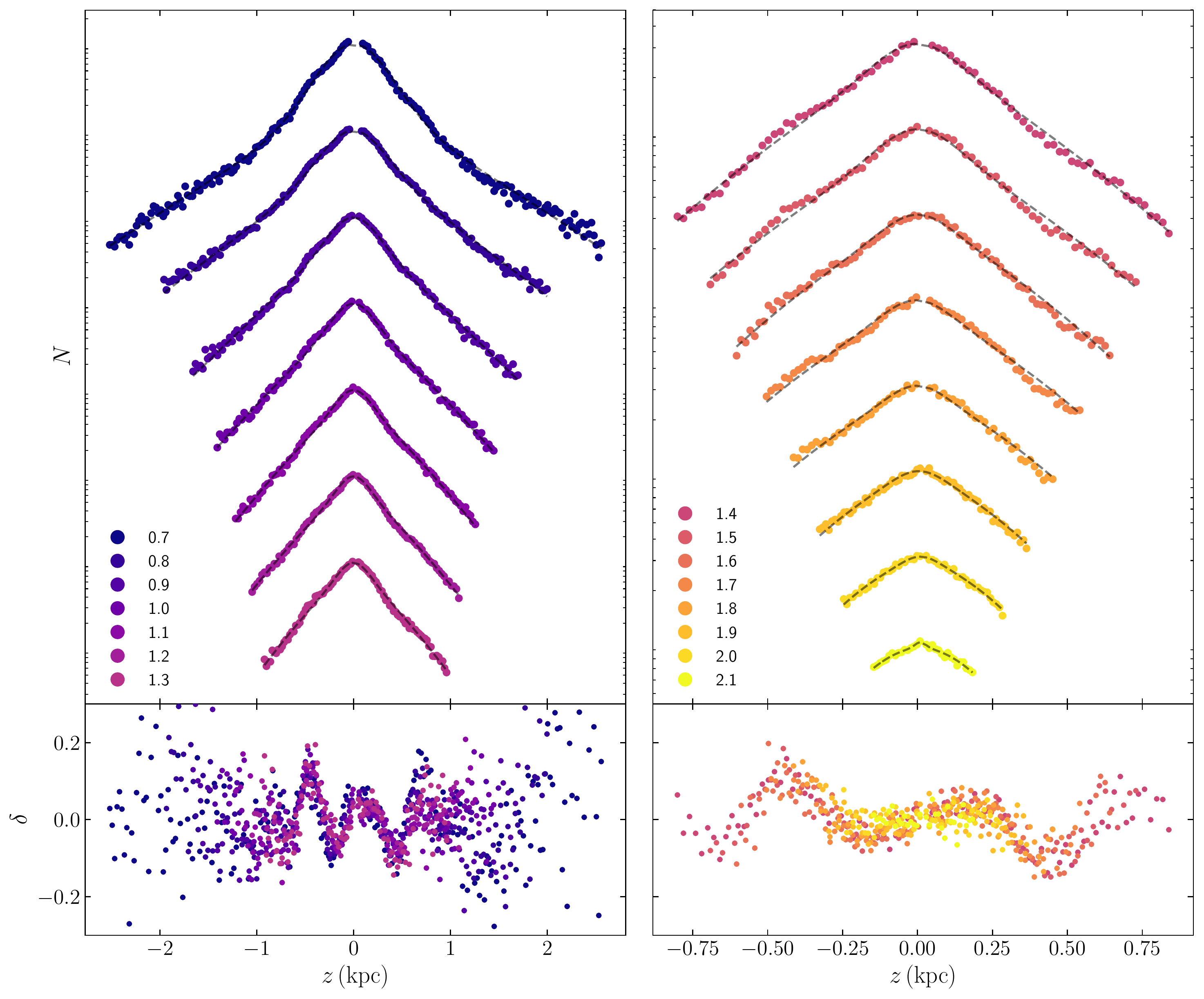}
    \caption{The raw number counts shifted by the average vertical position of the Sun, $20.3$ pc, so that zero is the Galactic mid-plane. The dashed lines are the best fit two-component models. The different colours represent different colour bins and are labeled by their associated \texttt{br\_min} value. The number counts have been artificially shifted along the y-axis to show trends as colour changes and not overlap. The bottom panel in both figures is the normalized residuals of the best fit. The residuals display a clear wave-like pattern that is the same in all colour bins.}
    \label{fig:ncount}
\end{figure*}
We first find the maximum-likelihood parameters using optimization and then we use this as an initial estimate for a Markov Chain Monte Carlo analysis. We perform this analysis for each colour bin separately, but in the end are mainly interested in obtaining a combined value for the solar position $z_\odot$, because this is what is used to correct the observed vertical heights. The uncertainties on the solar position are approximately equal for all colour bins, so we take the mean of all best fit values to represent the true position of the Sun. Through bootstrapping, we get a mean value of $z_\odot = 20.3\pm 0.7$ pc. 

The best fits as well as the raw number counts are shown in \figurename~\ref{fig:ncount}. Each colour bin is labeled by its \texttt{br\_min} value and the points are colour-coded such that the bluer bins are blue and redder bins are yellow. We also see a significant decrease in the range over which the sample is complete, between approximately 2.5 kpc for the bluest bin, to  around 0.2 kpc for the reddest bin. This behaviour is expected based on the associated magnitude ranges in \figurename~\ref{fig:cmd}. The bottom panel shows the normalized residuals, $\delta(z)=$ (N-model)/model. Much like \citet{widrow12}, they show an underlying oscillatory pattern to the number counts. 

These patterns are an indication of a departure from equilibrium in the Galactic disc. We can further characterize the perturbations to the disc by looking at the asymmetry parameter: 
\begin{equation}
A(z) = \frac{n(z)-n(-z)}{n(z)+n(-z)},
\label{eq:asym}
\end{equation}
where $n(z)$ is the true number count and not the equilibrium number count given by $N(z)$. To calculate this, we first adjust the vertical position of each star to account for the Sun's position, this ensures that the zero point is the mid-plane and not our location. Next, compute the number counts in bins in the corrected vertical position, and then use these raw number counts to calculate the asymmetry parameter. 

\begin{figure}
	\includegraphics[width=\columnwidth]{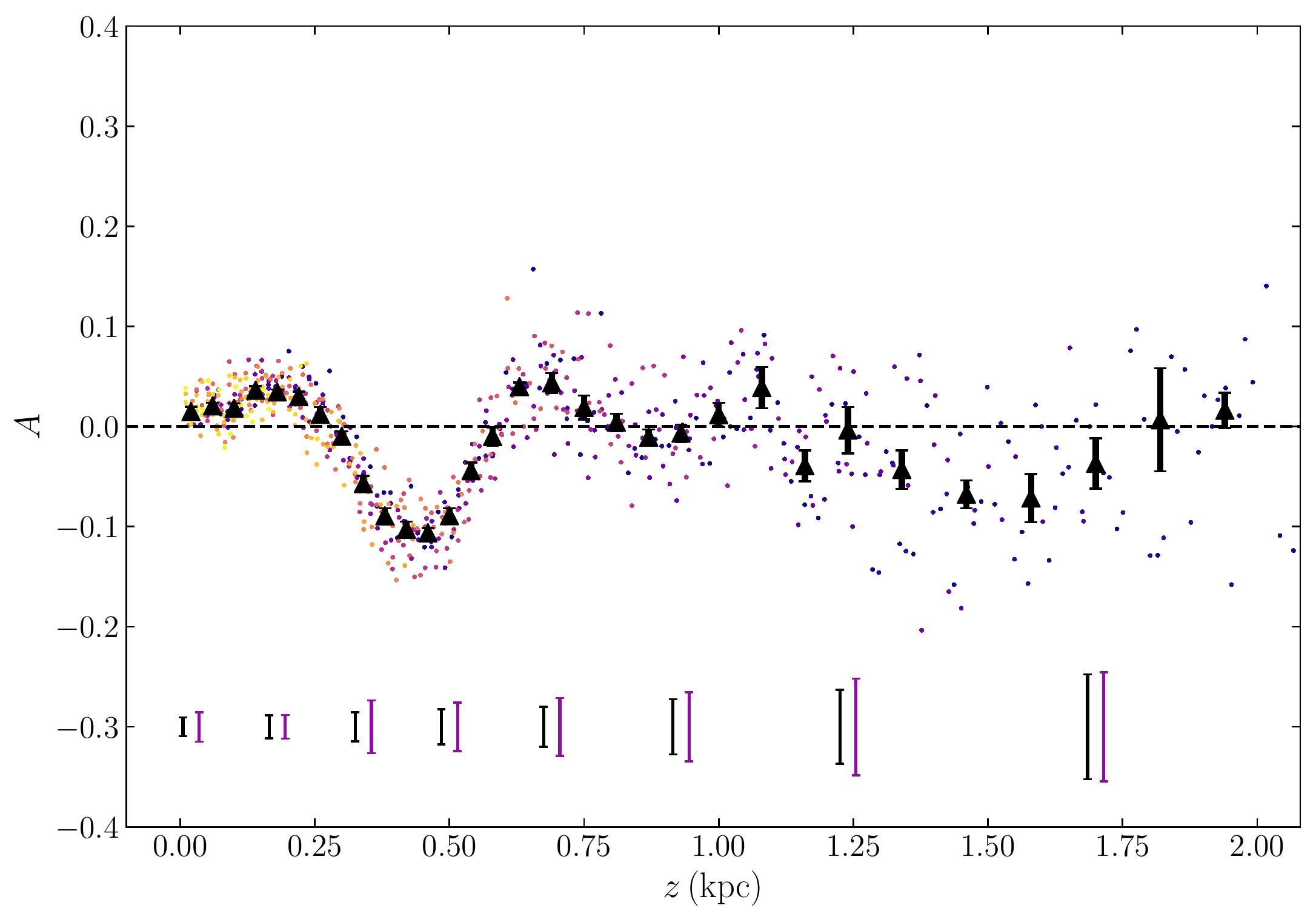}
    \caption{Asymmetry  in the vertical number counts as a function of vertical height. The smaller dots correspond to the different colour bins (colour coded as they were in \figurename~\ref{fig:ncount}). The typical error bar on each point is shown below (black) along with the spread of the points (purple). The larger triangles represent the median for different bins in vertical height. }
    \label{fig:asym}
\end{figure}

\subsection{Results}

The result of the analysis is shown in \figurename~\ref{fig:asym}. The colour bins follow the same colour coding used in \figurename~\ref{fig:ncount}. We then calculate the median and its uncertainties using a bootstrapping technique for different vertical heights across all the different colour bins. The median is plotted as the black triangles in \figurename~\ref{fig:asym}. The black error bar on the bottom of \figurename~\ref{fig:asym} shows the median of the uncertainties for each vertical bin and the purple error bar indicates the spread of the points. By comparing these two, it is evident that the spread in the asymmetry across colours can be explained by the errors in the points. 

The asymmetry clearly has an oscillatory behaviour. The trend also seems to be independent of colour bin which suggests it is a feature of the dynamics of the disc and therefore shares a common cause. Some of the notable features are the large dip just shy of 0.5 kpc and the peak at 0.7 kpc. This supports that the Galactic disc is undergoing some sort of wave perturbation as seen by previous works.  We discuss this new measurement of the asymmetry and how it relates to previous measurements in more detail in \secname~\ref{sec:zsun} below.

\section{Solar Height Above the Plane}\label{sec:zsun}

\begin{figure}	\includegraphics[width=\columnwidth]{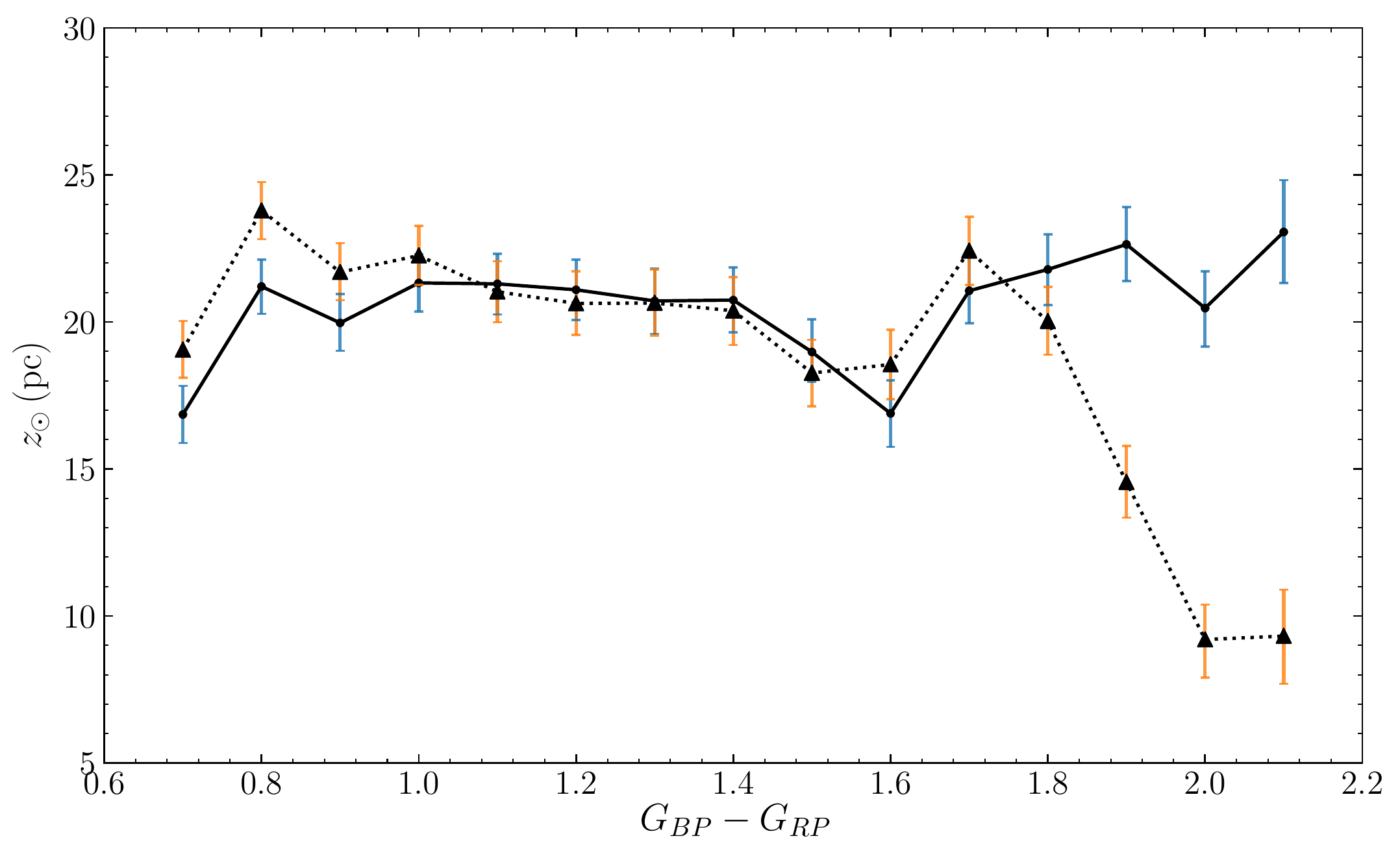}
    \caption{The vertical position of the Sun, $z_\odot$, as a function of the different colour bins. The colour bins are again labeled by their \texttt{br\_min} value. The triangles indicate the original best fit values and the points show the best fit when accounting for the asymmetry in the number counts. In both cases, uncertainties are calculated using an MCMC analysis. When accounting for the asymmetry, the redder colour bins give values for $z_\odot$ that are inconsistent with those of the bluer bins. Correcting for the asymmetry gives a consistent result for all colour bins and an overall best value of $z_\odot = 20.8 \pm 0.3$ pc.}
    \label{fig:zsun}
\end{figure}

As an example of how important considering the departure from equilibrium can be, we look at how including the asymmetry in the number counts affects the vertical position of the Sun with respect to the mid-plane. We define the local mid-plane as the peak in the symmetric component of the vertical stellar density. We can measure this from the observed number counts by accounting for the asymmetry that we determined above. However, this measurement of the vertical height is only correct up to an over-all bending mode of the disk. If there is a displacement of the entire disk, it would not be evident from our local analysis. In our original fits, we assume the asymmetry is small enough that it will not greatly affect the best fit parameters. However, when we examine the solar position for the different colours, there is a drop in the redder bins, as shown by the triangles in \figurename~\ref{fig:zsun}. These also happen to be the bins which are complete over a smaller height. Therefore, it is the most difficult to decouple the number count peak from waves due to asymmetry in these colour bins. 

The goal of this section is to see how accounting for the asymmetry in our model improves the solar height recovered from our best fit. This requires knowing the dependence of the true number count, $n(z)$, on the asymmetry parameter, $A(z)$, and the equilibrium number counts, $N(z)$. The number count asymmetry is anti-symmetric therefore, the underlying equilibrium number count can also be calculated via:
\begin{equation}
N(z)=\frac{n(z)+n(-z)}{2},
\end{equation}
We can then use this equation in combination with \eqnname~\eqref{eq:asym} to find the following relation:
\begin{equation}
n(z)= \begin{cases}
N(z)\left[1-A(z) \right ] & \text{ if } z\leq 0 \\ 
N(z)\left[1+A(z) \right ]  & \text{ if } z>0 
\end{cases}.
\label{eq:new_N}
\end{equation}
We estimate the form of the asymmetry by fitting a spline to the median asymmetry shown by triangles in \figurename~\ref{fig:asym}. Then we fit the observed number counts again using \eqnname~\ref{eq:new_N}, where $N(z)$ is given by \eqnname~\ref{eq:ncount}. By accounting for this additional behaviour in the number counts, the estimated solar height is improved and becomes $20.8\pm0.3$ pc. \figurename~\ref{fig:zsun} also shows that the trend at redder colours has been corrected by this new method. The points connected by a solid line show the solar position for different colour bins when accounting for the asymmetry, while the previous results are shown by the triangles and dashed line. At a colour of 1.7, the initial values of $z_\odot$ start to drop off. This has clearly been corrected when accounting for the asymmetry. We can therefore conclude, that even though the asymmetry is small, it is not always acceptable to ignore its effects on the system. 
Our measurement of the solar height above the Galactic mid-plane of $20.8\pm0.3$ pc agrees with values typical of previous measurements which vary between approximately 15 to 25 pc \citep[e.g.,][]{binneyzsun,chen01,Juric08}. The error on these types of measurements is typically on the order of a few parsecs, which we have greatly improved upon. However, some recent methods have improved upon this error margin. In particular, \citet{bovyselect} found a vertical height of $z_\odot= -0.9 \pm 0.9$. However, when he compares to measurements from subgiants, lower red giant branch, and red clump stars, the measured value is much larger. The discrepancy between the two groups could be explained by our analysis. The measurements of the number density profiles by \citet{bovyselect} extend, at most, out to 0.4 kpc and to even smaller heights for the earlier stellar types that give small $z_\odot$. Our analysis suggests that decoupling the height of the Sun above the Galactic mid-plane requires measurements out to greater heights if you do not account for the number count asymmetry. 

\section{Vertical Velocities}\label{sec:vel}

\begin{figure}
	\includegraphics[width=\columnwidth]{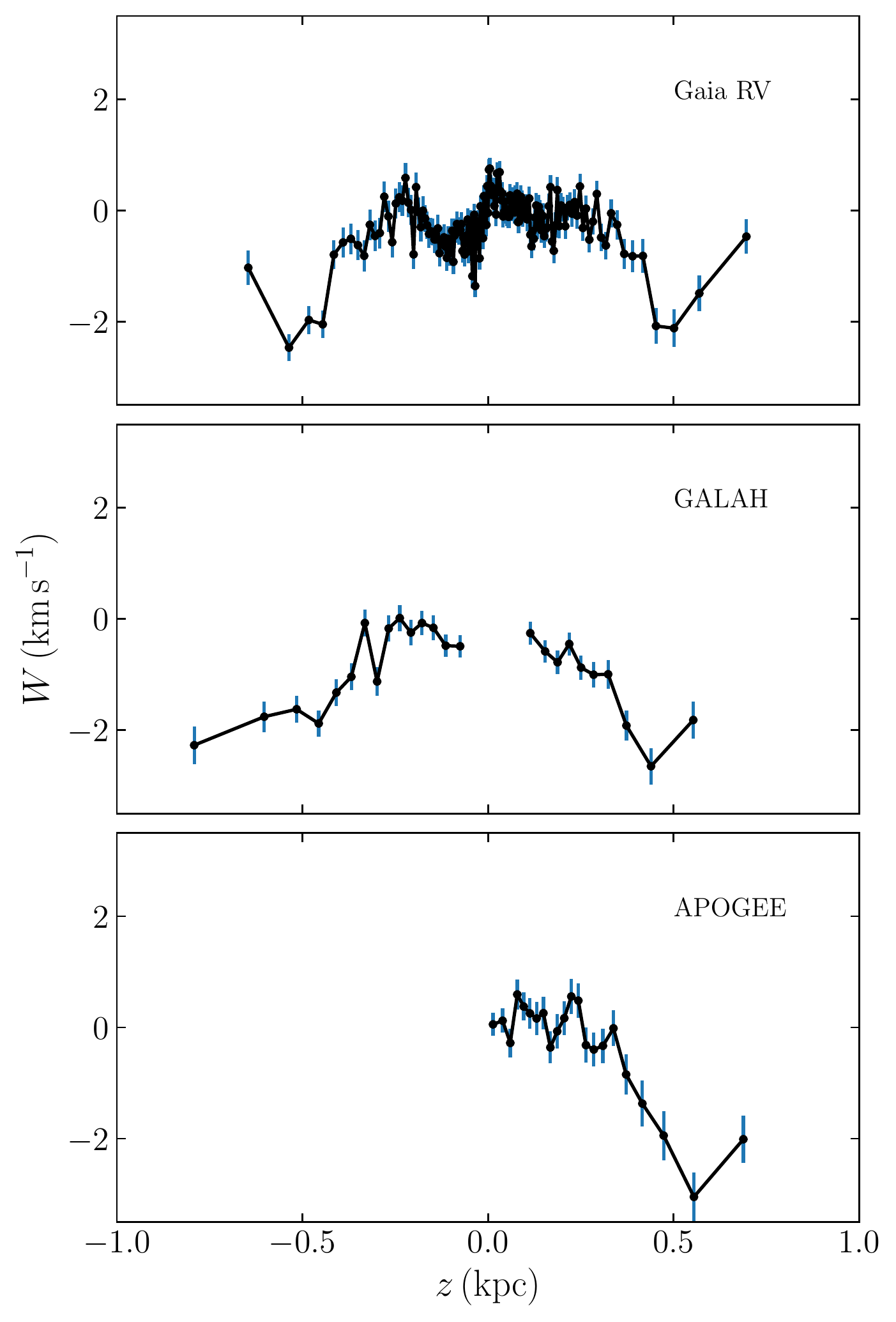}
    \caption{Running median of the vertical velocity as a function of vertical height. The three panels use different sets of data for the radial velocities: \gaia\ (top), GALAH (middle), and APOGEE (bottom). The samples are within a 2 kpc, 500 pc and 250 pc cylinder centered on the Sun, respectively.}
    \label{fig:3vel}
\end{figure}

Looking at just the number counts does not provide enough information about the Galactic disc to decouple the different perturbed modes. For an equilibrium system, the average vertical velocity should be zero across all heights, with as many stars traveling upwards as downwards. Any amplitude in the velocity indicates a departure from equilibrium. For this reason, we also look at the vertical velocities using the three surveys described in \secname~\ref{sec:data_vel}. There is a fairly large range of colours and magnitudes in these samples. However, the analysis of the number count asymmetry revealed that the perturbed behaviour is an intrinsic property of the disc and not just of a certain group of stars. This means that regardless of the stellar types, including giants, the behaviour is representative of the entire solar neighbourhood. A matching signal in the mean vertical velocity should therefore be present in the same form in dwarfs and giants.

To look at the trends in the vertical velocity, we use the radial velocity and proper motion in the radial ascension and declination. They are first transformed to the rectangular Galactic coordinate frame using the \texttt{galpy.util} Python package\footnote{\url{http://github.com/jobovy/galpy}} \citep{galpy}. We then calculate the running median of the sample ordered by vertical height for each survey and correct for the Sun's velocity by adding $7.25\,\textrm{km\,s}^{-1}$ \citep{vsun}. This works by calculating the median in a rolling window with a set size. This ensures that there are always enough points in each bin for the errors on the median to be small. The error in each point is estimated as approximately Gaussian. 

The distances for all samples are calculated using the \gaia\ DR2 parallaxes and are restricted to maximum 20\% error. For the GALAH sample, we look at stars within 2 kpc of the Sun so that we have enough data to look out to similar ranges to APOGEE and \gaia. This results in a sample with 160,686 stars. For the APOGEE data, we restrict the data set to within 0.5 kpc of the Sun which leaves us with 85,234 stars. Finally, with the \gaia\ radial-velocity data we restrict the sample to stars in a 250 pc cylinder around the Sun like we did for the number count samples.

\figurename~\ref{fig:3vel} shows the median velocities for \gaia, GALAH and APOGEE in order from top to bottom. For the APOGEE sample, we have restricted the plot to only the North, as there were not enough stars in the South to distinguish patterns. All three show evidence of a dip in velocities around 0.5 kpc above the mid-plane. Additionally, there appears to be a drop in velocity 0.5 kpc below in both the GALAH and \gaia\ samples. By and large, the trends seen in the different surveys are similar.

Furthermore, \gaia\ looks like it might have additional structure in the velocities closer to the mid-plane. To investigate this structure, we choose to calculate the median in bins of constant width as well as their uncertainties using a bootstrapping technique. This is shown in \figurename~\ref{fig:summary} where each bin is 25 pc within 0.5 kpc of the Sun and a width of 100 pc further out. The dips at $z=\pm 0.5$ kpc are consistent with those seen in the running mean.

\section{Discussion and Conclusions}\label{sec:dandc}

\begin{figure}
	\includegraphics[width=\columnwidth]{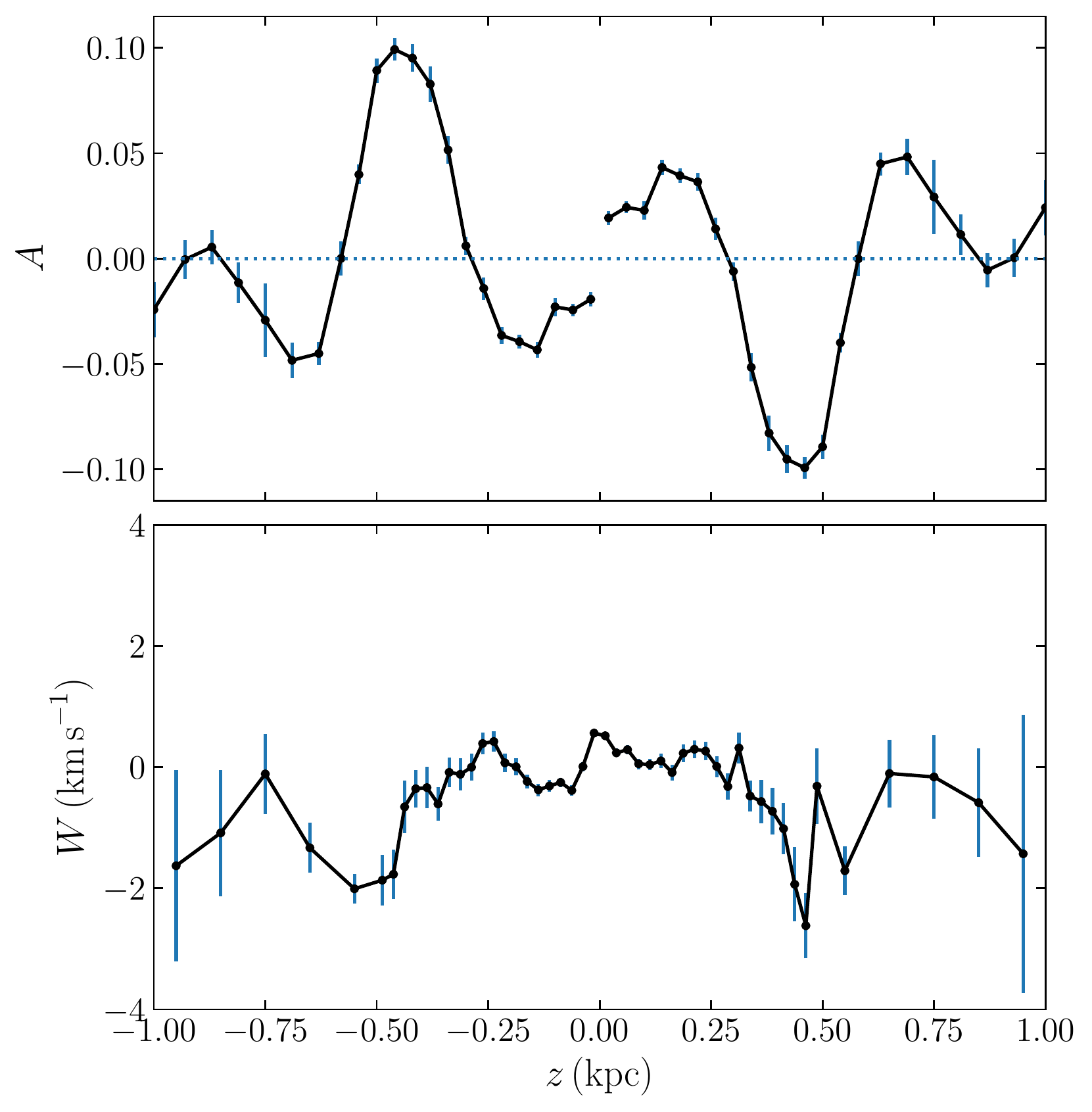}
    \caption{The top panel shows the number count asymmetry using $z_\odot= 20.8$ pc within a height of 1 kpc. It has been plotted for both positive and negative values to show the alignment with features in the velocity. The bottom panel shows the median vertical velocity using the \gaia\ DR2 radial velocity sample within a 250 pc radius cylinder centered on the Sun. The bins have a constant width of 25 pc for $|z|<500$ pc and constant width of 100 pc outside of that range. The number count asymmetry and the mean vertical velocity show trends that are plausibly dynamically consistent: the density asymmetry is accompanied by a symmetric velocity dip.}
    \label{fig:summary}
\end{figure}

Our goal in this paper was to use \gaia\ DR2 to assess the current state of the departure from vertical equilibrium in the Galactic disc using both the number count asymmetry and trends in the vertical velocity. The initial measurement of the asymmetry was made by \citet{widrow12} with only 300,000 main sequence stars. \citet{yanny13} were able to include a larger sample of stars, but still had to rely on spectroscopic parallax to recover distances. Finally, both of the previous analyses did not have access to a survey with full-sky coverage which complicates the selection function as you do not have equal volumes above and below the Galactic mid-plane. We are fortunate to be able to use \gaia\ DR2, which has the advantage of having geometric parallaxes, full-sky completeness, and a large number of stars and mitigates these areas of uncertainty in the analysis when compared to previous surveys. \figurename~\ref{fig:summary} shows the asymmetry using the new $z_\odot$ calculated in \secname~\ref{sec:zsun}. The dip in the number count asymmetry at approximately 0.4 kpc is consistent with previous works. Our study is able to probe closer to the galactic mid-plane than has been done before and therefore observe the continuation of the wave through the plane. While previous studies have found peaks at approximately 0.8 kpc, we uncover additional structure with double peaks at 0.7 kpc and 1.1 kpc. \citet{yanny13} also found the same gradual dip in the asymmetry further out than 1 kpc. Overall, our asymmetry measurements agrees with previous studies, but places these on a firmer basis and we find some possible additional structure.  

We further investigated trends in the vertical velocity above and below the Galactic mid-plane. We find a symmetric dip at approximately $\pm$0.5 kpc. This dip can also be seen in the measurement of the mean vertical velocity in Figure C.7 of \citet{gaia18a}. It is difficult to compare to other velocity measurements because the signal we see is approximately 2 km s$^{-1}$ at its largest value, and previous large surveys in the solar neighbourhood have not had the benefit of that level of accuracy. \citet{carrillo18} look at the vertical velocity trends over a range of radii, and similar trends can be seen in their $8<R<9$ kpc range. The large number of stars in \gaia\ DR2 allow us to examine unprecedented structure in the velocity, including even smaller fluctuations near the mid-plane. However, our velocity measurements do not extend as far vertically as previous studies.

We also compute the amplitude of the breathing mode as $v_\mathrm{breath}= 0.5\left[W(z)- W(-z)\right]$ using the mean vertical velocities $W(z)$ from \figurename~\ref{fig:summary} and find that it is consistently $<1\,\mathrm{km\,s}^{-1}$. In fact, the amplitude of the breathing mode is zero within the uncertainty at all heights. This is surprising considering that the first detection of oscillations in the Galactic disc concluded there was a breathing mode \citep{widrow12}. However, the results are consistent with observations made by \citet{gaia18a}. In their Figure C.6, they plot the amplitudes of the breathing mode for different heights and locations in the plane as well as the bending mode amplitudes. Especially close to the mid-plane, there appears to be a much stronger bending than breathing mode. We therefore conclude that there is no discernible breathing mode in the solar neighbourhood within 1 kpc of the mid-plane.

\figurename~\ref{fig:summary} summarizes our findings: we plot the density asymmetry in the top panel and the mean-velocity trend in the bottom panel. As pointed out by \citet{widrow12}, an asymmetry in the density cannot be caused by an overall bending of the local disc plane, which would simply offset the entire density distribution without changing its shape, or by a breathing mode, which would cause density perturbations symmetric around the mid-plane. An asymmetry in the density must be accompanied by a mean-velocity signal that is symmetric around the mid-plane. This is exactly what we observe in \figurename~\ref{fig:summary}: the mean vertical velocity has a symmetric dip on a similar scale, $\approx 0.5\,\mathrm{kpc}$, as the largest asymmetric feature in the density. Thus, we may for the first time be seeing the same local, vertical perturbation in both the density and vertical velocity.

The question now becomes, what can we learn from these observations? The coherence across the different colour bins means that stars of all ages are undergoing the same oscillatory motion. This suggests it was a singular dynamical event which excited the waves in the disc, a concept which is supported by other recent analyses \citep{widrow14,antoja18}. Future work modeling the impact of perturbations, such as those from a satellite fly-by, will benefit from having the detailed density and velocity perturbations that we present in \figurename~\ref{fig:summary} and, by using this information, may be able to unambiguously determine the properties of the perturber. Finally, accounting for the perturbations to the vertical dynamics will be important in any work attempting to measure the local disc and dark-matter density from modeling the local stellar dynamics. As a small first step in this direction, we have shown that the determination of the Sun's offset from the Galactic mid-plane is affected by the density asymmetry. Taking the asymmetry into account, we have been able to provide the most precise and accurate measurement of the Sun's position: $z_\odot = 20.8\pm 0.3\,\mathrm{pc}$.

\section*{Acknowledgements}

It is our pleasure to thank Jason Hunt for helpful comments. We would also like to acknowledge the anonymous referee for insightful and constructive comments. MB and JB received support from the Natural Sciences and Engineering Research Council of Canada (NSERC; funding reference number RGPIN-2015-05235) and from an Ontario Early Researcher Award (ER16-12-061). JB also received partial support from an Alfred P. Sloan Fellowship.

This work has made use of data from the European Space Agency (ESA) mission {\it Gaia} (\url{https://www.cosmos.esa.int/gaia}), processed by the {\it Gaia} Data Processing and Analysis Consortium (DPAC, \url{https://www.cosmos.esa.int/web/gaia/dpac/consortium}). Funding for the DPAC has been provided by national institutions, in particular the institutions participating in the {\it Gaia} Multilateral Agreement.






\bsp	
\label{lastpage}
\end{document}